\documentclass[%
 preprint,
nofootinbib,
 amsmath,amssymb,
 aps,
 pre,
]{revtex4}

\usepackage{bbm}
\usepackage{graphicx}
\usepackage{float}
\usepackage{dcolumn}
\usepackage{bm}
\usepackage{physics}
\usepackage{xcolor}
\usepackage{hyperref}

\numberwithin{equation}{section} 

\definecolor{red}{rgb}{1.00, 0.00, 0.00}

\begin{document}

\newcommand{\eins}{\leavevmode\hbox{\small1\kern-3.8pt\normalsize1}}
\newcommand{\hc}{^\dagger}
\newcommand{\nn}{\nonumber}
\newcommand{\inv}{^{-1}}
\newcommand{\be}{\begin{eqnarray}}
\newcommand{\ee}{\end{eqnarray}}

\title{Universal distributions from non-Hermitian Perturbation of Zero-Modes}%

\author{M. Kieburg}
\email{m.kieburg@unimelb.edu.au}
\affiliation{School of Mathematics and Statistics, University of Melbourne,\\
813 Swanston Street, Parkville, Melbourne VIC 3010, Australia}

\author{A. Mielke}%
\email{amielke@math.uni-bielefeld.de}
\affiliation{Faculty of Physics, Bielefeld University,\\
P.O. Box 100131, D-33501 Bielefeld, Germany}%

\author{M. Rud}
\email{zbh921@alumni.ku.dk}
\author{K. Splittorff}
\email{split@nbi.ku.dk}
\affiliation{Niels Bohr Institute, University of Copenhagen,\\
Blegdamsvej 17, 2100 Copenhagen, Denmark}%

\date{\today}

\begin{abstract}

Hermitian operators with exact zero modes subject to non-Hermitian perturbations are considered. Specific focus is on the average distribution of the initial zero modes of the Hermitian operators. The broadening of these zero modes is found to follow an elliptic Gaussian random matrix ensemble of fixed size, where the symmetry class of the perturbation determines the behaviour of the modes. This distribution follows from a central limit theorem of matrices, and is shown to be robust to deformations of the average.

\end{abstract}


\maketitle



\section{Introduction}

The microscopic spectrum and the zero modes of a Hermitian Hamiltonian matrix are of particular interest since these are usually closely linked to the symmetries and topology of the system. Studies of the smallest eigenvalues and zero modes therefore allow us to extract fundamental information about the system. This fact also extends to the Dirac operator in Quantum Chromodynamics (QCD) where the average spectral properties of the microscopic eigenvalues are intimately linked to  chiral symmetry breaking \cite{BC} and the zero modes by the Atiyah-Singer index theorem follow from the topology of the gauge field.  

Because of universality \cite{ADMN}, the average properties of the microscopic spectrum of the QCD Dirac operator may be studied either by random matrix theory  \cite{Verbaarschot:1994qf, ADMN, Mielke:2016kbg} or by means of effective field theory \cite{DOTV}. 
Similarly the smallest eigenvalues of the Hamiltonian for solid state systems carrying Majorana modes may be described either by random matrix theory or the associated $\sigma$-model \cite{BeenakkerRMT,BeenakkerMajorana}. 

In physical realizations, the symmetries and the topology of the ideal system is often perturbed, causing the zero modes to move slightly  away from the  origin. This poses a question: {\sl Is it possible in the presence of the perturbation to distinguish modes that have their origin in the topological zero modes of the unperturbed system?} This question is highly relevant for the study of Majorana modes \cite{BagretsAltland,HK,BeenakkerMajorana,Wilczek,Hamiltonian,ZKM,Neven,Dumitrescu} as well as in QCD \cite{DelDebbio:2005qa,KieburgWilson,MarioJacWilson,DSV,ADSV}.
A first answer was found in \cite{DSV,ADSV,MarioJacWilson} in the context of QCD where it was realized that indeed it is possible to distinguish would-be zero modes, since their distribution have very specific spectral statistics and scale differently with the volume than the rest of the small eigenvalues. Surprisingly, it was found that the spectral statistics of the would-be zero modes is given by finite size Gaussian random matrix theory. This at first came as a surprise as random matrix universality is usually only established in the limit of infinite matrix dimension. This new form of universality was shown \cite{Kieburg:2019gmn} to have its origin in a kind of matrix-valued central limit theorem that applies to the perturbation matrix for the zero modes. The new finite size Gaussian random matrix universality for the would-be zero modes is reached in the limit where the size of the remaining system is taken to infinity. In \cite{Kieburg:2019gmn}, Hermitian perturbations of the original ideal Hermitian system were considered, and it was possible to shown that universal distributions of perturbed zero modes exist for all universality classes. 

In the present work, we generalize this statement to weakly non-Hermitian perturbations. Systems with weak non-Hermiticity have been subject to several studies previously, both in the context of solid state physics \cite{Fyodorov:1996sx, Fyodorov:1997zz} as well as for the strong interactions \cite{Stephanov,Osborn,Splittorff:2006vj,AOSV,OSV}. 
We are in particular interested in non-Hermitian perturbations which violate the conditions responsible for the presence of exact zero modes in the original ideal system. The general matrix model we will consider is {of the form}
\begin{align}\label{eq:Model}
K=A+P,
\end{align}
where $A$ is a Hermitian matrix describing the ideal system with exact symmetries and exact zero modes. The non-Hermitian perturbation is given by $P$. 

We restrict our study to two classes of perturbations. In the first case, the perturbation $P_{\rm unc}$ takes the form 
\begin{align}\label{eq:Model_unc}
P_{\rm unc}=\alpha_{R}U_{R}S_{R}U_{R}^{\dagger}+ i \alpha_{I}U_{I}S_{I}U_{I}^{\dagger} \ ,
\end{align}
where {$S_{R}$ and $ i S_{I}$} are Hermitian and anti-Hermitian respectively and the real constants, $\alpha_{R}$ and $\alpha_{I}$, are chosen to be small such that the first order correction dominates. Their ratio will determine the ellipticity of the level density for the broadened zero eigenvalues.	 We will consider the average spectral properties of the smallest eigenvalues of $K$ where the average is over the unitary matrices $U_{R}$ and $U_{I}$. The average is taken over the respective Haar measure $d\mu(U)d\mu(V)$, where the two matrices $U_{R}$ and $U_{I}$ are considered to be statistically independent and there are therefore no correlations between the Hermitian and anti-Hermitian part.

The second class of perturbations $P_{\rm c}$ that we consider are of the form
\begin{align}\label{eq:Model_c}
P_{\rm c}=\alpha USV^{\dagger}.
\end{align}
The operator $S$ is fixed and complex-valued. The unitary matrices $U$ and $V$ are drawn from the deformed Haar measure
\be\label{eq:Model_c_Haar}
e^{z\Re\Tr\left[UV\hc\right]}d\mu(U) d\mu(V),
\ee
where $z>0$ is a fixed number setting the eccentricity of the limiting elliptic support of the level density. This freedom of alignment in the complex plane is missing in the first model where the spectrum is either elongated along the real or imaginary axis. Note that the Hermitian and the anti-Hermitian parts of this perturbation are now correlated. For $z\to\infty$ and fixed matrix size, the two unitary matrices become almost the same $U\approx V$.

We find that the finite size universality of the would-be zero modes extends to both types of non-Hermitian perturbations. The universal distributions are given by finite size non-Hermitian Gaussian elliptic random matrix ensembles. We also demonstrate numerically the validity of the universal finite size distributions for two specific random matrix realizations.

Because of the specific structure of the non-Hermitian perturbation in \eqref{eq:Model_unc} and \eqref{eq:Model_c}, we may take advantage of the symmetry classification of Hermitian matrices done by Altland and Zirnbauer \cite{Zirnbauer,Altland:1997zz}. Those non-Hermitian symmetric matrix spaces are classified in~\cite{LeClair,Magnea,Kawabata,Gong}. We will consider all those symmetry classes in unified way for both of our models. Surprisingly, both models yield the same limiting elliptic Gaussian ensemble whose joint matrix distribution have the same formal structure. They only differ in the symmetries of the matrices. Thus, we dub all of these ensembles elliptic Ginibre ensembles.

In Section \ref{sec:two_realizations} we introduce two physical realizations of the matrix model, which we will later analyse numerically and compare to our analytical findings. We continue in Section \ref{sec:char_eq} to show how the characteristic equation of the perturbation factorizes, such that the broadening of the zero modes is totally described by a specific subpart of the full perturbative term. In Section \ref{sec:dist} we show that the average spectral properties, of the perturbed zero modes, are determined by an elliptic Ginibre ensemble with size determined by the number of zero modes. In Section \ref{sec:numerical} we compare this analytical result with numerical analyses of the two ensembles, which were introduced in Section \ref{sec:two_realizations}. In Section \ref{sec:bounds} the bounds on the sizes of perturbation are studied, and the universality of the distribution of the former zero modes is considered. Finally, in Section \ref{sec:conclusion} we conclude and discuss our results.


\section{Symmetry Classes - Two Realizations}
\label{sec:two_realizations}

To motivate the study and exemplify the results, we consider two realizations of physical interest. A chiral ensemble, being of relevance to QCD, and an ensemble with particle-hole symmetry relevant for solid-state systems with Majorana zero modes.

\subsection{Chiral Ensemble}
The QCD Dirac operator for vanishing quark masses exhibits a chiral symmetry and the spontaneous breaking hereof induces a non-zero eigenvalue density of the Dirac operator at the origin \cite{BC}. Furthermore the winding number of the gauge field gives rise to eigenvalues exactly at the origin, see eg.~\cite{Leutwyler:1992yt}. However, when simulating QCD on the lattice in the Wilson approach, the influence of the non-zero lattice spacing breaks both the chiral symmetry and also perturbs the zero modes. This affects the distribution of the smallest eigenvalues of the Dirac operator in a highly non trivial manner \cite{DSV,ADSV,MarioJacWilson}. When a chemical potential is introduced, we break Hermiticity \cite{AOSV,OSV} as well. It therefore becomes interesting to study the effect of a chiral ensemble perturbed by a non-Hermitian one.

This motivates us to study the following general non-Hermitian perturbation of the chiral random matrix model
\begin{align}
	K^{(N)}=
    	\begin{pmatrix}
      	0 & M \\
      	M^{\dagger} & 0
    	\end{pmatrix}
    	+\alpha_{R}U_{R}S_{R}U_{R}^{\dagger}+ i \alpha_{I}U_{I}S_{I} U_{I}^{\dagger},
\end{align}
where the initial Hamiltonian has dimensions $N\times N$ with $N=2n+\nu$, and $M$ has dimensions $(n+\nu)\times n$. Thus, for $\alpha_{R}=\alpha_{I}=0$ the model exhibits $\nu$ zero modes. This allows us to control the exact number of zero modes, we wish to study. The matrices $S_{R}$ and $S_{I}$ have dimensions $N\times N$, are Hermitian and have otherwise no further symmetries.  The only average is performed over the unitary matrices $U_{R}$ and $U_{I}$, which are drawn via the Haar measure from the group determined by the symmetries of the ensemble. The real coupling constants $\alpha_{R}$ and $\alpha_{I}$ control the magnitude of the Hermitian and the anti-Hermitian part of the perturbation, respectively.

We will show that the perturbation matrix for the former zero modes is distributed according to a Gaussian in the complex plane with standard deviations $\sigma_l=\alpha_l\sqrt{\Tr (S_l^{(j_l)})^2/(\widetilde{\gamma}_{l}^{(j)}(N_{l}^{(j)})^2)}$ where $l=R,I$ is the real and imaginary direction respectively. The parameters $\widetilde{\gamma}_{l}^{(j)}$ depend on the symmetry class. See Eq.~(\ref{widetildegamma}) below. It is basically the exponent of the determinant when performing a multivariate Gaussian integral, see table II in~\cite{Kieburg:2019gmn}. It follows that the broadening of the zero modes is given by the one-point correlation function of the elliptic Ginibre ensemble \cite{akemann_baik_francesco}
\begin{align}
	R_{1}(\lambda)=\sqrt{\omega_{\tau}(\lambda)\omega_{\tau}				(\lambda^{*})}\sum_{n=0}^{\nu-1}\frac{\tau^{n}}{2^{n}n!}H_{n}			\left(\frac{\lambda}{\sqrt{2\tau}}\right)H_{n}\left(\frac{\lambda^{*}}			{\sqrt{2\tau}}\right),
    	\label{one_point_corr}
\end{align}
where
\begin{align}
\omega_{\tau}(\lambda) = \frac{1}{\pi\sqrt{1-\tau^2}} \exp(-\frac{|\lambda|^2}{1-\tau^2} + \frac{\tau(\lambda^2 + {\lambda^*}^2)}{2(1-\tau^2)})
\end{align}
and the factor $\tau$ is defined as
\begin{align}
	\frac{\sqrt{1+\tau}}{\sqrt{1-\tau}}=\frac{\alpha_{R}						\sqrt{\Tr\left(S_{R}\right)^{2}}}{\alpha_{I} \sqrt{\Tr\left(S_{I}\right)^{2}}}.
\end{align}

\subsection{Majorana Ensemble}

In the study of superconductors carrying topological zero modes, see for instance \cite{BeenakkerMajorana,Kitaev,vonOppen,FlensbergReview,Neven,Elliot,Dumitrescu,Chiu}, it becomes relevant to study the spectral properties of a Hamiltonian consisting of the direct sum of two antisymmetric Hermitian ensembles (Class D in the Cartan classification) that are the same up to a sign. This corresponds to particle-hole-symmetry \cite{BeenakkerRMT,Neven,Chiu}, and the antisymmetric matrices are of odd size, which gives the system two identical zero modes. It has been pointed out that it is difficult to experimentally distinguish such an ensemble where the zero modes have been perturbed from an ensemble with an accumulation of eigenvalues around the origin \cite{BagretsAltland}. It is therefore advantageous to study perturbations arising from thermal fluctuations that couple the two sectors. We here consider a generalization where the perturbation is non-Hermitian. Such a non-Hermiticity may arise from a coupling to an environment. This is usually the case in scattering systems.

In terms of a random matrix model this ensemble is:
\begin{align}
  	K^{(N)}=
    	\begin{pmatrix}
      	 i M & 0\\
      	0 & - i M
    	\end{pmatrix}
    	+\alpha_{R}O_{R}
    	\begin{pmatrix}
      	0 &  i W_{R}\\
      	- i W_{R}^{T} & 0
    	\end{pmatrix}
    	O_{R}^{T}+ i \alpha_{I}O_{I}
    	\begin{pmatrix}
      	0 &  i W_{I}\\
      	- i W_{I}^{T} & 0
    	\end{pmatrix}
    	O_{I}^{T}, && M=-M^{T},
\end{align}
where $M$ is real antisymmetric and has dimensions $2n+\nu$. $W_{R}$ and $W_{I}$ also have dimensions $2n+\nu$, are real but have no further symmetries. This ensemble is defined for all $\nu$, but we need only $\nu=0,1.$ The average is done over the orthogonal matrices $O_{R}$ and $O_{I}$ drawn from the Haar measure of the orthogonal group. As above $\alpha_{R}$ and $\alpha_{I}$ are real and control the magnitude of the Hermitian and anti-Hermitian part of the perturbation, respectively.

As mentioned above, we will show that the perturbation matrix of the zero modes has a Gaussian distribution in the complex plane with standard deviations $\sigma_l=\alpha_l\sqrt{\Tr (S_l^{(j_l)})^2/(\widetilde{\gamma}_{l}^{(j)}(N_{l}^{(j)})^2)}$ where $l=R,I$. This ensemble exhibits two exact zero modes when $\alpha_{R}=\alpha_{I}=0$, hence, for $\nu$ odd the perturbation matrix is of dimension $(4n+2\nu)\times (4n+2\nu)$ and therefore $N=4n+2\nu$. The full ensemble matrix $K^{(N)}$ is antisymmetric, $K^{T}=-K$, which makes the perturbation matrix antisymmetric as well.
If the perturbation matrix is distributed according to a Gaussian, then the broadened zero modes will be as well, see the ensuing example with a $2\times2$ antisymmetric matrix:
\begin{align}
	0=\det\left[ i \begin{pmatrix}
		-\lambda & x\\
		-x & -\lambda
	\end{pmatrix}\right]
	\Rightarrow \lambda^{2}-x^{2}=0.
\end{align}
The probability density for elements with Gaussian weights is defined as
\begin{align}
	P(\lambda)=\int_{-\infty}^{\infty}dx\: \delta (\lambda^{2}-x^{2})e^{-x^{2}/\sigma^{2}}=e^{-\lambda^{2}/\sigma^{2}}.
\end{align}
Therefore, the broadened zero modes are compared to Gaussian distributions in both the real and imaginary direction with the shown standard deviations,
\begin{align}
    	p(\lambda)=\frac{\exp\left(-\left(\Re [\lambda]\right)^{2}/\sigma_R^2-\left(\Im [\lambda]\right)^{2}/\sigma_I^2\right)}{\pi\sigma_R\sigma_I}.
\label{Majorana_gauss}
\end{align}
Note that our result for the model \eqref{eq:Model_unc} hold for all non-Hermitian classes. This is clear because we will show that our results hold for $S_R$ and $S_I$ in any Hermitian symmetry class. This will certainly give an over-counting, but we are guaranteed to cover all classes with non-correlated Hermitian and anti-Hermitian part.

For the model \eqref{eq:Model_c}, this is less clear, primarily because the non-Hermitian matrices have not yet been unambiguously classified \cite{LeClair,Magnea,Kawabata,Gong}. We still conjecture that they hold in full generality, because symmetry constraints will merely give conditions on $U$ and $V$. It is likely that the integrals with substructure may be solved the same way as for the different Hermitian classes.


\section{Factorization of the Characteristic Equation}
\label{sec:char_eq}

First we show that, for sufficiently weak perturbation, the spectrum of the former zero modes decouples from the bulk. This corresponds to focusing on the part of the perturbation which corresponds to first order degenerate perturbation theory. The approach here follows closely the one in \cite{Kieburg:2019gmn}.
We may deal with both forms \eqref{eq:Model_unc} and \eqref{eq:Model_c} in a unified fashion.
Our starting point is the following
\begin{align}\label{eq:SimplifiedPerturb}
T = \alpha U S V\hc
\end{align}
with
\begin{align}\label{eq:defU}
U = \left(\begin{array}{c}
U_1\\
U_2
\end{array}\right) ,\ V = \left(\begin{array}{c}
V_1\\
V_2
\end{array}\right)
\end{align}
unitary and divide it into the sectors
\begin{align}
T = \left(\begin{array}{c|c}
T_1 & T_2\\
\hline
T_4 & T_3
\end{array}\right)\ .
\end{align}
We have organized the basis such that $T_3$ will correspond to the zero modes of $A$ that is given in this basis as follows
\begin{align}
A= \left(\begin{array}{c|c}
A' & 0\\
\hline
0 & 0
\end{array}\right)\ .
\end{align}
The notation is chosen in such a way to line up with $S_3$, $S_{R3}$, and $S_{I3}$, as in~\cite{Kieburg:2019gmn}.

Although the form~\eqref{eq:SimplifiedPerturb} resembles  the second model~\eqref{eq:Model_c} instead of the first one~\eqref{eq:Model_unc}, both can be dealt here in the same way. One needs to keep in mind that the result in \cite{Kieburg:2019gmn} gives a bound on $\alpha\norm{S}_{\rm op}$, where $\norm{\cdot}_{\rm op}$ is the operator norm (the largest singular value). As we shall see, the corresponding bound here will depend on $\norm{\alpha_{R}U_{R}S_{R}U_{R}^{\dagger}+ i \alpha_{I}U_{I}S_{I}U_{I}^{\dagger}}_{\rm op}$ or $\norm{ \alpha U S V\hc}_{\rm op}=\alpha\norm{ S}_{\rm op}$. We underline that the operator norm is invariant under left and right multiplication of unitary matrices. Thus, when defining $\tilde{U}=U_{R}^{\dagger}U_{I}$ we can identify $S=\alpha_{R}S_{R}+ i \alpha_{I}\tilde{U}S_{I}\tilde{U}^{\dagger}$. Since $\tilde{U}$ is Haar distributed, we can have any combination of the eigenvalues of the real eigenvalues of $\alpha_{R}S_{R}$ and the imaginary eigenvalues of $ i \alpha_{I}S_{I}$. Therefore, we have the inequality $\norm{\alpha_{R}S_{R}+ i \alpha_{I}\tilde{U}S_{I}\tilde{U}^{\dagger}}_{\rm op}\leq \sqrt{\alpha_R^2\norm{S_R}_{\rm op}^2+\alpha_I^2\norm{S_I}_{\rm op}^2}$. The equality is given by the worst case when $S_R$, $S_I$ and $\tilde{U}$ are diagonal and the eigenspaces of the largest eigenvalues of $S_R$ and $S_I$ are the same. In practice, this will not be the case, and the inequality will hold instead. Hence, we need to replace the operator norm by the norm $\sqrt{\alpha_R^2\norm{S_R}_{\rm op}^2+\alpha_I^2\norm{S_I}_{\rm op}^2}$ for the estimates in the first model.

We consider the characteristic polynomial
\begin{align}\label{eq:eig1}
\det(K^{(N)}-\lambda \eins_N) &=\det\left(\begin{array}{c|c}
A' + T_1-\lambda\eins_{N-\nu} & T_2\\
\hline
T_4 & T_3-\lambda\eins_\nu
\end{array}\right)\\
&=\det\left(A' - \lambda\eins_{N-\nu}\right) \det\left(\eins_{N-\nu} +  (A'-\lambda\eins_{N-\nu})\inv T_1 \right)\nonumber\\
&\times \det\left[T_3 - \lambda\eins_\nu - T_4\left(\eins_{N-\nu} + \left(A' - \lambda \eins_{N-\nu}\right)\inv T_1 \right)\inv \left(A' - \lambda\eins_{N-\nu}\right)\inv T_2 \right],\nonumber
\end{align}
where we first pulled out the  factor $A' - \lambda\eins_{N-\nu}$ and then the factor $\eins_{N-\nu} +  (A'-\lambda\eins_{N-\nu})\inv T_1$. Next, we use $T_1=U_1 \alpha S V_1\hc$ and express the inverse as a Neumann series
\begin{align}
\left(\eins_{N-\nu} + \left(A' - \lambda \eins_{N-\nu}\right)\inv U_1 \alpha S V_1\hc \right)\inv = \sum_{j=0}^{\infty} \left[-\left(A' - \lambda \eins_{N-\nu}\right)\inv U_1 \alpha S V_1\hc\right]^j,\nonumber
\end{align}
which can be exploited as follows
\begin{align}
& U_2 \alpha S \left(\eins_{N-\nu} - V_1\hc\left(\eins_{N-\nu} + \left(A' - \lambda \eins_{N-\nu}\right)\inv U_1  \alpha S V_1\hc \right)\inv \left(A' - \lambda\eins_{N-\nu}\right)\inv U_1  \alpha S V_2\hc\right)\nonumber\\
=& U_2 \alpha S \left(\eins_{N-\nu} +  \sum_{j=1}^{\infty} \left[-V_1\hc\left(A' - \lambda \eins_{N-\nu}\right)\inv U_1  \alpha S \right]^j\right)V_2\hc\nonumber\\
=& U_2  \alpha S \left[\eins_{N-\nu} + V_1\hc\left(A' - \lambda \eins_{N-\nu}\right)\inv U_1  \alpha S \right]\inv V_2\hc\ .\nonumber
\end{align}
Here, we have made use of $T_2=U_1 \alpha SV_2^\dagger$ and $T_4=U_2 \alpha SV_1^\dagger$.
Inserting this result in Equation \eqref{eq:eig1}, the characteristic polynomial becomes
\begin{align}\label{eq:eig2}
\det(K^{(N)}-\lambda \eins_N)=& \det\left(A' - \lambda\eins_{N-\nu}\right)\det\left(\eins_N + \alpha S V_1^\dagger(A'-\lambda\eins_{N-\nu})\inv U_1\right)\\
&\times\det\left(U_2[\eins_N + \alpha S V_1^\dagger(A'-\lambda\eins_{N-\nu})\inv U_1]\inv \alpha S U_2^\dagger -\lambda \eins_\nu\right),\nonumber
\end{align}
which is in full analogy to the Hermitian case in~\cite{Kieburg:2019gmn}.

The same analogy carries even further. To make the first order perturbation theory exact we need the smallest eigenvalues of $A'$, given by $\norm{(A')^{-1}}_{\rm op}^{-1}$, to not interact with the broadened spectrum of the zero modes. The latter is represented by the operator in the third derterminant of~\eqref{eq:eig2}. Hence, we need
\begin{equation}\label{eq:ConditionFactor.b}
\begin{split}
\norm{(A')^{-1}}_{\rm op}^{-1}\gg& \norm{U_2[\eins_N + \alpha S V_1^\dagger(A'-\lambda\eins_{N-\nu})\inv U_1]\inv \alpha S U_2^\dagger}_{\rm op}\\
=&\alpha\norm{[\eins_N + \alpha S V_1^\dagger(A'-\lambda\eins_{N-\nu})\inv U_1]\inv S }_{\rm op}.
\end{split}
\end{equation}
We underline that $\lambda$ is of the order of the operator on the right hand side so that we can drop it in every combination of the form $A'-\lambda\eins_{N-\nu}$. This simplifies the characteristic polynomial to

\begin{align}
\det(K^{(N)}-\lambda \eins_N)\approx& \det\left(A'\right)\det\left(\eins_N + \alpha SV_1^\dagger(A')\inv U_1\right)\nn\\
&\times\det\left(U_2[\eins_N +  \alpha S V_1^\dagger(A')\inv U_1]^{-1} \alpha S V_2^\dagger -\lambda \eins_\nu\right) .
\end{align}
Furthermore, we would like to suppress the operator $\alpha S V_1^\dagger(A')\inv U_1$ compared to the identity matrix $\eins_N$. In particular we need $\norm{\alpha S V_1^\dagger(A')\inv}_{\rm op}\ll 1$. To understand of what order the generic value of $\norm{\alpha S V_1^\dagger(A')\inv}_{\rm op}$ is, we choose an arbitrary vector $\ket{\chi}\in\mathbf{C}^N$ and consider the $m$'th moments of the squared norm of $V_1^\dagger(A')\inv U_1  \alpha S \ket{\chi}$, which gives the conditions
\begin{align}\label{eq:ConditionFactor}
\begin{split}
\int_{\mathcal{K}} d\mu(U)(\bra{\chi} \alpha S\hc U_1^\dagger(A')^{-2} U_1  \alpha S \ket{\chi})^m \leq & c_m \left(\frac{\alpha^2\Tr (A')^{-2} \bra{\chi} S\hc S\ket{\chi}}{N}\right)^m \ll 1 .
\end{split}
\end{align}
Here $c_m$ is some constant of order 1 and $\mathcal{K}$ is the Haar measure of the appropriate unitary group. 
Since the vector $\ket{\chi}$ is arbitrary, we can also choose the eigenvector to the largest singular value $\norm{S}_{\rm op}$ of $S$.
Combining Eqs.~\eqref{eq:ConditionFactor.b} and~\eqref{eq:ConditionFactor}, we need to assume
\begin{align}\label{eq:ConditionFactor_cor}
\alpha\frac{\sqrt{\Tr (A')^{-2}} \norm{S}_{\rm op}}{\sqrt{N}} \ll 1 
\end{align}
for the second model, which indeed also implies $||A^{-1}||_{\rm op}^{-1}\gg\alpha||S||_{\rm op}$, and
\begin{align}\label{eq:ConditionFactor_uncor}
\frac{\sqrt{\Tr (A')^{-2}} \norm{\alpha_{R}U_{R}S_{R}U_{R}^{\dagger}+ i \alpha_{I}U_{I}S_{I}U_{I}^{\dagger}}_{\rm op}}{\sqrt{N}}\leq \sqrt{\frac{\Tr (A')^{-2}( \alpha_R^2\norm{S_R}_{\rm op}^2+\alpha_I^2\norm{S_I}_{\rm op}^2)}{N}}\ll 1 
\end{align}
for the first model. These two conditions allow us to restrict the discussion to first order perturbation theory. In practice this means that we only need to study the spectrum of
\begin{align}
\alpha_RS_{R3}+i\alpha_I \ S_{I3} = \alpha_RU_{R2} S_{R} U_{R2}\hc+i\alpha_IU_{I2} S_{I} U_{I2}\hc \quad {\rm or}\quad \alpha S_3=\alpha U_2SV_2^\dagger
\end{align}
as these matrices determine the distribution of the former zero modes. We will derive the distribution of these matrices in the next section.

Before coming to calculating this distribution, we would like to emphasise the following. Although first order perturbation theory of the first model may dictate that the spectrum of a purely real or imaginary perturbation only spreads in the according direction, eventually the spectrum will start to invade the whole complex plane. 


\section{Distribution of $S_{3}$ - a matrix-valued central limit theorem}
\label{sec:dist}

We move on with a derivation of the distribution of the perturbation matrix and how this broadens the zero modes of our Hamiltonian. We consider the two classes \eqref{eq:Model_unc} and \eqref{eq:Model_c} in turn.

\subsection{Perturbation with uncorrelated Hermitian and anti-Hermitian parts}

We start by considering perturbations of the form $P_{\rm unc}$ of \eqref{eq:Model_unc}. Since the Hermitian and anti-Hermitian parts are uncorrelated the result from the Hermitian case \cite{Kieburg:2019gmn}  can be exploited for the two parts independently. To apply this result we need the additional conditions
\begin{align}\label{ratio-limit}
    \lim_{N\to\infty}\frac{\sqrt{\Tr\left(S_{R/I}\right)^{2}}}{\norm{S_{R/I}}_{op}}=\infty
\end{align}
and
\begin{align}
    \Tr\left(S_{R/I}\right)&=0.
    \label{eq:traceless}
\end{align}
These essentially state that a big portion of the singular values have to be of the same order as the largest one. Looking into the proof of the Hermitian case in~\cite{Kieburg:2019gmn}, this condition guarantees that the distributions of $S_{R3}$ and $S_{I3}$ converge to Gaussians. Without it, deviations from the Gaussian behavior are quite likely.

For the sake of generality, we choose $S_R$ and $S_I$ being independently a direct sum of operators in one of the ten symmetry classes rather than just a single one. The direct sum is highly important because some ensembles in the Magnea--Bernard--LeClair classification~\cite{LeClair,Magnea,Kawabata} have real and imaginary parts that decompose in direct sums, e.g., the Wilson--Dirac operator~\cite{DSV,ADSV,MarioJacWilson,KieburgWilson}. This happens exactly then when $S$ has a pseudo-Hermiticity property, meaning $S^\dagger=\gamma_5S\gamma_5$ with $\gamma_5=\gamma_5^\dagger=\gamma_5^{-1}$. Then there is a basis where the Hermitian part is block diagonal and the anti-Hermitian part is chiral, meaning the direct sum has maximal two components in the standard symmetry classification by Magnea~\cite{Magnea}. The role of Hermitian and anti-Hermitian part may be reversed when having $S^\dagger=-\gamma_5S\gamma_5$. The notation of $\gamma_5$ is reminiscent of the $\gamma_5$ matrix in the four dimensional Dirac theory, like the QCD Dirac operator.

The dimension of the subspaces for the decompositions $S_R=\bigoplus_{j_R} S_R^{(j_R)}$ and $S_I=\bigoplus_{j_I} S_I^{(j_I)}$ are $N_{R}^{(j_R)}$ and $N_{I}^{(j_I)}$, respectively. The corresponding unitary matrices $U_R$ and $U_I$ have an according decomposition $U_R=\bigoplus_{j_R} U_R^{(j_R)}$ and $U_I=\bigoplus_{j_I} U_I^{(j_I)}$, where $S_R^{(j_R)}\to U_R^{(j_R)}S_R^{(j_R)}(U_R^{(j_R)})^\dagger$ and $S_I^{(j_I)}\to U_I^{(j_I)}S_I^{(j_I)}(U_I^{(j_I)})^\dagger$ keeps the respective global symmetries invariant and generates the largest compact groups distributed along the corresponding Haar measure. Then, the distribution of each matrix $S'_{l,j}=U_l^{(j)}S_l^{(j)}(U_j^{(j)})^\dagger$ with $l=R,I$ is for large $N$ given by~\cite{Kieburg:2019gmn}
\begin{equation}
p(S'_{l,j})=\frac{\exp\left[-\widetilde{\gamma}_{l}^{(j)}(N_{l}^{(j)})^2\Tr (S'_{l,j})^2/\Tr (S_I^{(j_I)})^2\right]}{\int d\tilde{S}\exp\left[-\widetilde{\gamma}_{l}^{(j)}(N_{l}^{(j)})^2\Tr \tilde{S}^2/\Tr (S_I^{(j_I)})^2\right]}
\end{equation}
with the denominator properly normalizing the distribution and $\widetilde{\gamma}_{l}^{(j)}$ a parameter of order one that depends on the chosen symmetry class of the Altland--Zirnbauer classification~\cite{Zirnbauer,Altland:1997zz}, i.e.,
\begin{equation}\label{widetildegamma}
\widetilde{\gamma}_{l}^{(j)}=\left\{\begin{array}{cl}
\displaystyle \frac{1}{4}, & \text{non-chiral and an anti-unitary symmetry},\\
1, & \text{non-chiral and no anti-unitary symmetry or}\\
& \text{one of the two Boguliubov--de Gennes operators},\\
\displaystyle\frac{p_{l}^{(j)}n_{l}^{(j)}}{(N_{l}^{(j)})^2}, & \text{chiral and no  anti-unitary symmetry},\\
\displaystyle\frac{p_{l}^{(j)}n_{l}^{(j)}}{2(N_{l}^{(j)})^2}, & \text{the remaining two chiral operator clases}.\\
\end{array}\right.
\end{equation}
For the standard chiral ensembles we have the operator dimensions $p_{l}^{(j)},n_{l}^{(j)}\propto N_{l}^{(j)}\gg1$ with $p_{l}^{(j)}+n_{l}^{(j)}=N_{l}^{(j)}$.
To be precise, we need to assume $\lim_{N\to\infty}N_l^{(j)}=\infty$, $\Tr S_{l}^{(j)}=0$ as well as  $\lim_{N\to\infty}\sqrt{\Tr\left(S_{l}^{(j)}\right)^{2}}/\norm{S_{l}^{(j)}}_{op}=\infty$ for each $l=R,I$ and $j$ to find this result. The traceless condition is a non-trivial restriction for a decomposition into a true direct sum. This has to be seen in contrast to when no direct sum is present. Here the whole operator is only shifted by a scalar times the identity matrix. This is not true for a direct sum, where we get a shift by a diagonal matrix consisting of blocks of different scalars in front of the identity matrix when we do not have the trace condition. This has indeed physical effects as known from the Wilson--Dirac random matrix model~\cite{DSV,ADSV,MarioJacWilson,KieburgWilson} and its corresponding lattice QCD-operator~\cite{Wilson}.

In total, the distribution of the full perturbation $S'_3=\alpha_RS_{R3}+i\alpha_I \ S_{I3}=\alpha_R\bigoplus_{j_R} S_R^{(j_R)}+i\alpha_I\bigoplus_{j_I} S_I^{(j_I)}$  converges in the large $N$-limit to the joint Gaussian probability density
\begin{align}\label{eq:GaussianS3}
    p(S'_{3})=\prod_{l=R,I}\prod_{j_l} p(S'_{l,j_l}).
\end{align}
This constitutes the matrix central limit theorem for the perturbation matrices $S_3'$ mentioned in the introduction.
When assuming that for both $l=R,I$ there is a $\sigma_l=\alpha_l\sqrt{\Tr (S_I^{(j_I)})^2/(\widetilde{\gamma}_{l}^{(j)}(N_{l}^{(j)})^2)}$ for all $j$, then we have the simplification of the distribution to the elliptic Ginibre ensemble for one of the non-Hermitian ensembles~\cite{LeClair,LeClair}
\begin{align}\label{eq:GaussianS3.b}
    p(S'_{3})=\frac{\exp[-\Tr\left(S_{R3}\right)^{2}/\sigma_R^2-\Tr\left(S_{I3}\right)^{2}/\sigma_I^2]}{\int d\tilde{S}_{R3}d\tilde{S}_{I3}\exp[-\Tr\left(\tilde{S}_{R3}\right)^{2}/\sigma_R^2-\Tr\left(\tilde{S}_{I3}\right)^{2}/\sigma_I^2]}.
\end{align}

Let us summarize and underline once again that the homogeneity $\sigma_l=\alpha_l\sqrt{\frac{\Tr (S_I^{(j_I)})^2}{\widetilde{\gamma}_{l}^{(j)}(N_{l}^{(j)})^2}}$ and being traceless $\Tr S_{l}^{(j)}=0$ are for the single components of a direct sum of operators not always guaranteed, e.g., for the Wilson--Dirac operator~\cite{DSV,ADSV,MarioJacWilson,KieburgWilson}. Thus, the Gaussian distribution becomes non-centered and has a non-trivial covariance matrix. Nevertheless, the joint distribution of $S'_3=\alpha_RU_{R2} S_{R} U_{R2}\hc+i\alpha_IU_{I2} S_{I} U_{I2}\hc$ remains Gaussian as long as the conditions $\lim_{N\to\infty}\sqrt{\Tr\left(S_{l}^{(j)}\right)^{2}}/\norm{S_{l}^{(j)}}_{op}=\infty$ for all $l=R,I$ and $j$ as well as Eq.~\eqref{eq:ConditionFactor_uncor} are satisfied.

\subsection{Perturbation with correlated Hermitian and anti-Hermitian parts and deformed Haar measure}

We now turn to the second form where the perturbation is given by \eqref{eq:Model_c}. The condition that corresponds to~\eqref{ratio-limit} reads
\begin{eqnarray}\label{ratio-limit2}
\lim_{N\to\infty} q^{(N)} = \infty
\end{eqnarray}
where
\begin{eqnarray}
	q^{(N)}=\frac{\sqrt{\Tr SS^\dagger}}{\norm{S}_{\rm op}} .
\end{eqnarray}
We again also assume $\Tr S=0$ to simplify the problem.  As introduced in Eqs.~\eqref{eq:Model_c} and \eqref{eq:Model_c_Haar}, we consider the perturbation
\begin{eqnarray}
K &=& A + \alpha US V\hc.
\end{eqnarray}
This time the perturbation does not need to decompose into a direct sum since we always consider the sum of the Hermitian and anti-Hermitian part as a whole. This detail can be implemented in $S=e^{i\varphi}(\cos(\vartheta)S_R+i\sin(\vartheta)S_I)$ where $S_R$ and $S_I$ are the two Hermitian components in one of the non-Hermitian symmetry classes~\cite{Magnea} and $\vartheta,\varphi\in[-\pi,\pi]$ embed the perturbation somehow in the complex matrices. Since the phase $e^{i\varphi}$ is only a scalar rotation, we can set it to $1$ for the computation and later multiply it to the matrix again.

To  understand the nature of the unitary matrices $U$ and $V$, we need to consider different cases that concerns the global symmetries. One symmetry is a possible pseudo-Hermiticity like $S^\dagger=\gamma_5S\gamma_5$. Here we have the relation $V=\gamma_5U\gamma_5$. In the case of pseudo-symmetry like $S^T=\pm\gamma_5S\gamma_5$, this relation reads $V=\widetilde{\gamma}_5U^*\widetilde{\gamma}_5$ with a different $\widetilde{\gamma}_5=\widetilde{\gamma}_5^\dagger=\widetilde{\gamma}_5^{-1}$. Certainly, $S$ may fulfill none or both. In the latter situation $S$ as well as $U$ satisfy reality conditions, $S^*=\pm \gamma_5\widetilde{\gamma}_5S\widetilde{\gamma}_5\gamma_5$ and $U^*=\gamma_5\widetilde{\gamma}_5S\widetilde{\gamma}_5\gamma_5$.

To simplify the situation and not to discuss all four cases, separately, we exploit the fact that operators which do not satisfy one or both of these symmetries can be embedded always in those where this is indeed the case. To this aim we consider instead of $K=A + \alpha US V\hc$ the enlarged matrices,
\begin{equation}\label{embedding}
\begin{split}
{\rm diag}(K,K^*,K^\dagger,K^T)=&{\rm diag}(A,A^*,A^\dagger,A^T)\\
+&\alpha{\rm diag}(U,U^*,V,V^*){\rm diag}(S,S^*,S^\dagger,S^T)\widehat{\gamma}_5^{(N)}{\rm diag}(U^\dagger,U^T,V^\dagger,V^T)\widehat{\gamma}_5
\end{split}
\end{equation}
with
\begin{equation}\label{gamma.def}
\widehat{\gamma}_5^{(N)}=\left(\begin{array}{cc} 0 & \eins_{2N} \\ \eins_{2N} & 0 \end{array}\right).
\end{equation}
We relabel the dimension $4N=\hat{N}$, ${\rm diag}(K,K^*,K^\dagger,K^T)=\hat{K}$, ${\rm diag}(S,S^*,S^\dagger,S^T)=\hat{S}$, ${\rm diag}(U,U^*,V,V^*)=\hat{U}$ and so forth. We can therefore, without restriction of generality, assume that $S$ or better $\hat{U}\hat{S}\widehat{\gamma}_5^{(N)}\hat{U}^\dagger\widehat{\gamma}_5^{(N)}$ satisfies a pseudo-Hermiticity condition with the matrix $\widehat{\gamma}_5$ and a pseudo-symmetry condition with the matrix
\begin{equation}
\widetilde{\gamma}_5^{(N)}=\left(\begin{array}{c|c} 0 &\begin{array}{cc} 0 & \eins_{N} \\ \eins_{N} & 0 \end{array}  \\ \hline  \begin{array}{cc} 0 & \eins_{N} \\ \eins_{N} & 0 \end{array} & 0 \end{array}\right),
\end{equation}
and, therefore, also a reality condition which may lead to one of the three number fields. Indeed, $S$ may additionally satisfy commutation relations with fixed matrices. This is usually the case when it has a block-diagonal or chiral structure or when it has had already a pseudo-Hermiticity and/or pseudo-symmetry condition before the embedding~\eqref{embedding}.
Moreover, the unitary matrix $\hat{U}$ might be real, complex or quaternion keeping the symmetry under the complex conjugation as well as block-diagonal or a full matrix that keeps the possible chiral or block structures invariant. Those choices still reflect the Altland--Zirnbauer classification~\cite{Zirnbauer,Altland:1997zz}.

Once this is settled, we can go over to the distribution of the unitary matrix $U$, which is
\begin{eqnarray}\label{measure}
d\tilde{\mu}(\hat{U})=\frac{\exp\left(z\Tr\left[\hat{U}\widehat{\gamma}_5^{(N)}\hat{U}\hc\widehat{\gamma}_5^{(N)}\right]/4\right)d\mu(\hat{U})}{\int \exp\left(z\Tr\left[\hat{U}\widehat{\gamma}_5^{(N)}\hat{U}\hc\widehat{\gamma}_5^{(N)}\right]/4\right)d\mu(\hat{U})}\ {\rm with}\ z>0.
\end{eqnarray}
In the case when neither a pseudo-Hermiticity or pseudo-symmetry originally existed, this measure agrees with Eq.~\eqref{eq:Model_c_Haar} after  employing the embedding~\eqref{embedding}.
In any case, the deformation of the Haar measure in~\eqref{measure} has the effect that for $z\to\infty$, the unitary matrix starts to align to the relation $\hat{U}=\widehat{\gamma}_5^{(N)}\hat{U}\widehat{\gamma}_5^{(N)}$. Employing the original model with the embedding~\eqref{embedding}, this would translate to $U=V$. When $z$ is correlated with the angle $\vartheta$ such that $\vartheta\to0$ or $\vartheta\to\pi/2$ for $z\to\infty$, the perturbation becomes essentially a Hermitian matrix. In comparison, the limit $z\to0$ yields the unmodified Haar measure which always corresponds to the full non-Hermitian case. Since we have now two parameters, $z$ and $\tau$, which have the same impact, we can eliminated one of those. Therefore, we choose $S$ to be in one of the ten symmetry classes of the Altland--Zirnbauer scheme~\cite{Zirnbauer,Altland:1997zz}, meaning $\vartheta=0,\pi/2$. Then, we select one non-Hermitian symmetric space in the Magnea scheme~\cite{Magnea} where $S$ is either the Hermitian or anti-Hermitian part. Afterwards, we employ the embedding~\eqref{embedding}. The unitary matrix space for $\hat{U}$ is then the maximally compact subgroup of the invariance group of these symmetric spaces.

Let us give three examples to illustrate this construction:
\begin{enumerate}
\item		The first one is the choice of $S$ being a real symmetric matrix and embedded in the complex symmetric matrices. The invariance group of the complex symmetric matrices is given by the map $P\to GPG^T$ with $G\in{\rm Gl}(N,\mathbb{C})$ and, therefore, the general linear complex group. The corresponding maximally compact subgroup are the unitary matrices  with the map $P\to UPU^T$. Thus, $V=U^*$ and $U$ is a unitary matrix. The enlarged unitary matrix is $\hat{U}={\rm diag}(U,U^*,U^*,U)$. In the limit for $z\to\infty$, $U$ becomes a real orthogonal matrix.
\item For the second example we again choose $S$ to be real symmetric but now embedded in the real matrices. The invariance group of the real matrices is given by the map $P\to G_1PG_2^T$ with $(G_1,G_2)\in{\rm Gl}(N,\mathbb{R})\times{\rm Gl}(N,\mathbb{R})$. This time the maximally compact subgroup is the product of two real orthogonal groups ${\rm O}(N)\times{\rm O}(N)$  with the map $P\to UPV^T$ with the enlarged unitary matrix $\hat{U}={\rm diag}(U,U,V,V)$. When we take $z\to\infty$, the two matrices reduce to a single one via $U=V$.
\item		The last example is  given by the choice of $S$ being the direct sum of two Hermitian matrices of dimensions $n\times n$ and $(n+\nu)\times(n+\nu)$ embedded in the $\gamma_5={\rm diag}(\eins_n,-\eins_{n+\nu})$-Hermitian matrices, i.e., $P^\dagger=\gamma_5P\gamma_5$. The invariance group of these pseudo-Hermitian matrices is determined by the map $P\to GP\gamma_5G^\dagger\gamma_5$ with $G\in{\rm Gl}(N,\mathbb{C})$ and its maximally compact subgroup is  the unitary group ${\rm U}(N)$. Thence, the matrix $U$ is unitary and $V=\gamma_5U\gamma_5$ and $\hat{U}={\rm diag}(U,U^*,\gamma_5U\gamma_5,\gamma_5U^*\gamma_5)$. In this case, the limit $z\to\infty$ leads to a block-diagonal form of $U=\gamma_5U\gamma_5$.
\end{enumerate}

The derivation of the large $N$ limit follows closely the one for the Hermitian case, see~\cite{Kieburg:2019gmn}. The normalized Haar measure $d\mu(\hat{U})$ for $\hat{U}$ over the compact group $\mathcal{K}$ can be expressed in terms of a Dirac delta function in the matrix group $\mathcal{G}\supset \mathcal{K}$ which keeps the non-Hermitian matrices invariant, i.e.,
\begin{equation}
\int_{\mathcal{K}} f(\hat{U})d\mu(\hat{U})=\frac{\int_{\mathcal{G}} f(\hat{U})\delta(\eins_{\hat{N}}-\hat{U}\hat{U}^\dagger)d\hat{U}}{\int_{\mathcal{G}} \delta(\eins_{\hat{N}}-\hat{U}\hat{U}^\dagger)d\hat{U}}
\end{equation}
for an arbitrary suitably integrable function $f$. As we have seen in the three examples and can be readily checked in general, the closure of these matrix groups are  always vector spaces, in particular some kind of embeddings of Cartesian products of one of the three general linear groups. The Dirac delta function may, on the other hand, be expressed as a Fourier--Laplace transform on the Hermitian matrix space $\mathcal{H}={\rm span}(\mathcal{G}\mathcal{G}^\dagger)={\rm span}\{GG^\dagger| G\in\mathcal{G}\}$, where ${\rm span}$ is the linear span, i.e.,
\begin{equation}\label{eq:Haardelta}
\int_{\mathcal{K}} f(\hat{U})d\mu(\hat{U})=\lim_{\epsilon\to0}\frac{\int_{\mathcal{G}}d\hat{U}\int_{\mathcal{H}}d\hat{Q} f(\hat{U})\exp[\epsilon\Tr(\eins_{\hat{N}}-i\hat{Q})^2 +\xi\Tr(\eins_{\hat{N}}-\hat{U}\hat{U}^\dagger)(\eins_{\hat{N}}-i\hat{Q})]}{\int_{\mathcal{G}}d\hat{U}\int_{\mathcal{H}}d\hat{Q} \exp[\epsilon\Tr(\eins_{\hat{N}}-i\hat{Q})^2 +\xi\Tr(\eins_{\hat{N}}-\hat{U}\hat{U}^\dagger)(\eins_{\hat{N}}-i\hat{Q})]}.
\end{equation}
The matrix $\hat{Q}$ has the form ${\rm diag}(Q_1,Q_1^*,Q_2,Q_2^*)$ with $Q_1$ and $Q_2$  Hermitian matrices which may satisfy some relations if $U$ and $V$ do.
The auxiliary variable $\epsilon>0$ renders the integrals absolutely integrable and the parameter $\xi>0$ will be chosen appropriately when performing the saddle point analysis. The variable $\xi$ is independent of $\epsilon$ but depends on $N$, $z$ as well as the chosen symmetry class as we will see later on.

The function $f$ in the denominator  is the deformation of the Haar measure $e^{z\Tr[\hat{U}\widehat{\gamma}_5^{(N)}\hat{U}^\dagger\widehat{\gamma}_5^{(N)}]/4}$ and in the numerator we have additionally the Dirac delta function
\begin{equation}\label{eq:S3delta}
\begin{split}
&\delta(\hat{S}'_3-\kappa \hat{U}_2\hat{S} \widehat{\gamma}_5^{(N)}\hat{U}_2^\dagger\widehat{\gamma}_5^{(\nu)})=\lim_{\epsilon\to0}\frac{\int_{\mathcal{M}} dH \exp[-\epsilon\Tr (\hat{H}\widehat{\gamma}_5^{(\nu)})^2+i\Tr \hat{H}(\hat{S}'_3-\kappa \hat{U}_2\hat{S}\widehat{\gamma}_5^{(N)}\hat{U}_2^\dagger\widehat{\gamma}_5^{(\nu)})]}{\int_{\mathcal{M}}dS'_3\int_{\mathcal{M}} dH \exp[-\Tr (\hat{H}\widehat{\gamma}_5^{(\nu)})^2+i\Tr \hat{H}\hat{S}'_3]}
\end{split}
\end{equation}
with $\kappa>0$ again a parameter that has to be appropriately chosen. The matrix $\hat{H}$ depends on $H$ like $\hat{H}={\rm diag}(H,H^*,H^\dagger,H^T)$ and the same holds for $\hat{S}'_3$ and $S'_3$. Already in the first model we have seen that it is proportional to $N/\sqrt{\Tr SS^\dagger}$ which is also true here. The matrix space $\mathcal{M}$ is one of the symmetric non-Hermitian matrix spaces given in Magnea's work~\cite{Magnea} (note that with the embedding all classes have pseudo-Hermiticity). We would like to point out that we need the matrix $\widehat{\gamma}_5^{(j)}$ in two different dimensions, see~\eqref{gamma.def}, namely $j=N$ for the symmetry of $S$ and $j=\nu$ for the projected subspace of the broadened zero modes.

We write $U_2=\Pi_2U$ with $\Pi_2$ the projection onto the last $\nu$ rows and denote the embedded matrices by $\hat{U}_2$, $\hat{\Pi}_2$ and $\hat{U}$. Plugging Eq.~\eqref{eq:S3delta} into~\eqref{eq:Haardelta}, the integral over $\hat{U}$ becomes Gaussian. As already mentioned, $\hat{U}$ might be in one of the three number fields as well as have a block-diagonal structure that certainly contains $U$ and $U^*$ but never $U^T$ or $U^\dagger$ (this follows from the group multiplication property) as we know from the embedding. Thus, $\hat{U}\in\mathcal{G}$ can be understood as a very large real vector whose real entries may appear multiple times.
The multiplicity can only be $1,2,4,8$ when restricting $P=\alpha USV^\dagger$ to one of the standard non-Hermitian symmetry classes~\cite{LeClair,Magnea}.

Let $d_{\mathcal{G}}$ be the dimension of $\mathcal{G}$. Then, we have the following Gaussian integral
\begin{equation}\label{Gauss-int}
\begin{split}
&\frac{\int_{\mathcal{G}}d\hat{U} \exp[\frac{z}{4}\Tr[\hat{U}\widehat{\gamma}_5^{(N)}\hat{U}^\dagger\widehat{\gamma}_5^{(N)}]-i\kappa\Tr \hat{H}\hat{\Pi}_2 \hat{U}\hat{S}\widehat{\gamma}_5^{(N)}\hat{U}^\dagger\hat{\Pi}_2^\dagger\widehat{\gamma}_5^{(\nu)}-\xi\Tr\hat{U}\hat{U}^\dagger(\eins_{\hat{N}}-i\hat{Q})]}{\int_{\mathcal{G}}d\hat{U}\int_{\mathcal{H}}d\hat{Q} \exp[\frac{z}{4}\Tr[\hat{U}\widehat{\gamma}_5^{(N)}\hat{U}^\dagger\widehat{\gamma}_5^{(N)}]+\epsilon\Tr(\eins_{\hat{N}}-i\hat{Q})^2 +\xi\Tr(\eins_{\hat{N}}-\hat{U}\hat{U}^\dagger)(\eins_{\hat{N}}-i\hat{Q})]}\\
=&\frac{{\det}^{-d_{\mathcal{G}}/(2\hat{N}^2)}\left[\eins_{\hat{N}}\otimes(\eins_{\hat{N}}-i\hat{Q})+i\frac{\kappa}{\xi}\hat{\Pi}_2^\dagger\widehat{\gamma}_5^{(\nu)}\hat{H}\hat{\Pi}_2\otimes \hat{S}\widehat{\gamma}_5^{(N)}-\frac{z}{4\xi}\widehat{\gamma}_5^{(N)}\otimes\widehat{\gamma}_5^{(N)}\right]}{\int_{\mathcal{H}}d\hat{Q} \det^{-d_{\mathcal{G}}/(4\hat{N})}\left[(\eins_{\hat{N}}-i\hat{Q})^2-\frac{z^2}{16\xi^2}\eins_{\hat{N}}\right]\exp[\epsilon\Tr(\eins_{\hat{N}}-i\hat{Q})^2 +\xi\Tr(\eins_{\hat{N}}-i\hat{Q})]}.
\end{split}
\end{equation}
We exploited here that half of the eigenvalues of $\widehat{\gamma}_5^{(N)}$ are $+1$ and the other half is $-1$.

The saddle point analysis of the $\hat{Q}$ integral for large $N$ is dominated by the Lagrangian
\begin{equation}
L(\hat{Q})=\xi\Tr(\eins_{\hat{N}}-i\hat{Q})-\frac{d_{\mathcal{G}}}{4\hat{N}}\Tr\,{\rm ln}\left[(\eins_{\hat{N}}-i\hat{Q})^2-\frac{z^2}{16\xi^2}\eins_{\hat{N}}\right].
\end{equation}
The $\epsilon$ dependent term is tiny because $\epsilon$ will be sent to $0$. Moreover, we can read off the scaling of $\xi$ which has to be linear in $N$. We note that $d_{\mathcal{G}}=\mathcal{O}(N^2)$.
Due to the singularity of the integrand at $(\eins_{\hat{N}}-i\hat{Q})^2=z^2/(16\xi^2)$ preventing us from shifting the contour to other saddle points, we have a unique solution at
\begin{equation}
\eins_{\hat{N}}-i\hat{Q}_0=\frac{d_{\mathcal{G}}+\sqrt{d_{\mathcal{G}}^2+z^2\hat{N}^2}}{4\hat{N}\xi}\eins_{\hat{N}}=q\eins_{\hat{N}}
\end{equation}
When expanding the integrand in $\hat{Q}=\hat{Q}_0+\delta \hat{Q}/\sqrt{N}$, the integral over the massive modes $\delta \hat{Q}$ becomes a Gaussian and can be carried out. When choosing $\kappa$ appropriately as we will do soon, the remaining determinant in the  numerator will not change this saddle point approximation. We are left with
\begin{equation}
\begin{split}
p(S'_3)=&\lim_{\epsilon\to0}\frac{\int_{\mathcal{M}} dH \exp[-\epsilon\Tr (\hat{H}\widehat{\gamma}_5^{(\nu)})^2+i\Tr \hat{H}\hat{S}'_3+\tilde{L}(H)]}{\int_{\mathcal{M}}dS'_3\int_{\mathcal{M}} dH \exp[-\Tr (\hat{H}\widehat{\gamma}_5^{(\nu)})^2+i\Tr \hat{H}\hat{S}'_3]}
\end{split}
\end{equation}
with
\begin{equation}
\begin{split}
\tilde{L}(H)=&-\frac{d_{\mathcal{G}}}{2\hat{N}^2}\Tr\,{\rm ln}\left[q\eins_{\hat{N}^2}+i\frac{\kappa}{\xi}\hat{\Pi}_2^\dagger\widehat{\gamma}_5^{(\nu)}\hat{H}\hat{\Pi}_2\otimes \hat{S}\widehat{\gamma}_5^{(N)}-\frac{z}{4\xi}\widehat{\gamma}_5^{(N)}\otimes\widehat{\gamma}_5^{(N)}\right]+\frac{d_{\mathcal{G}}}{4}{\rm ln}\left(q^2-\frac{z^2}{16\xi^2}\right)\\
=&-\frac{d_{\mathcal{G}}}{2\hat{N}^2}\Tr\,{\rm ln}\left[\eins_{\hat{N}^2}+i\frac{4\kappa}{16\xi^2q^2-z^2}\hat{\Pi}_2^\dagger\widehat{\gamma}_5^{(\nu)}\hat{H}\hat{\Pi}_2\otimes \hat{S}\widehat{\gamma}_5^{(N)}\left(4\xi q\eins_{\hat{N}^2}-z\widehat{\gamma}_5^{(N)}\otimes\widehat{\gamma}_5^{(N)}\right)\right].
\end{split}
\end{equation}
When choosing $\kappa$ of the order $\mathcal{O}(1/\sqrt{\Tr SS^\dagger})$, we can expand the logarithm in $\hat{S}$. The first term vanishes exactly because $\Tr \hat{S}=4 {\rm Re}\Tr S=0$ and $\Tr \hat{S}\widehat{\gamma}_5^{(N)}=0$. The higher order terms asymptotically vanish in the large $N$ limit because of the estimate
\begin{equation}
\left|\frac{\Tr [\hat{S}(c_1\eins_{\hat{N}}+c_2\widehat{\gamma}_5^{(N)})]^j}{(\Tr SS^\dagger)^{j/2}}\right|\leq4(|c_1|+|c_2|)^j\left(\frac{\norm{S}_{\rm op}}{\sqrt{\Tr SS^\dagger}}\right)^{j-2}=4(|c_1|+|c_2|)^j\left(q^{(N)}\right)^{2-j}\overset{N\to\infty}{\to}0
\end{equation}
with two constants $c_1$ and $c_2$ of order one or smaller, as it is the case for us. For the limit we exploited the assumption~\eqref{ratio-limit2}.

Summarizing the Lagrangian $\tilde{L}(H)$ asymptotes to the quadratic term
\begin{equation}
\tilde{L}(H)\overset{N\gg1}{\approx}-\frac{4d_{\mathcal{G}}\kappa^2}{\hat{N}^2(16\xi^2q^2-z^2)^2}\biggl[ 16\xi^2q^2\Tr(\widehat{\gamma}_5^{(\nu)}\hat{H})^2\ \Tr(\hat{S}\widehat{\gamma}_5^{(N)})^2+z^2 \Tr \hat{H}^2\ \Tr\hat{S}^2\biggl].
\end{equation}
The mixed term vanishes because $\Tr \hat{S}^2\widehat{\gamma}_5^{(N)}=0$, and we have employed $\hat{\Pi}_2\widehat{\gamma}_5^{(N)}\hat{\Pi}_2^\dagger=\widehat{\gamma}_5^{(\nu)}$. Since we now have a integration-guaranteeing term we can perform the limit $\epsilon\to0$, exactly. Hence, we end up with the distribution
\begin{equation}\label{intermediate}
\begin{split}
p(S'_3)=&\frac{\int_{\mathcal{M}} dH \exp[i\Tr \hat{H}\hat{S}'_3+\tilde{L}(H)]}{\int_{\mathcal{M}}dS'_3\int_{\mathcal{M}} dH \exp[-\Tr (\hat{H}\widehat{\gamma}_5^{(\nu)})^2+i\Tr \hat{H}\hat{S}'_3]}\\
=&\frac{\exp[-\tilde{a}\Tr (\widehat{\gamma}_5^{(\nu)}\hat{S}'_3)^2+\tilde{b}\Tr (\hat{S}'_3)^2]}{\int_{\mathcal{M}}dS'_3 \exp[-\tilde{a}\Tr (\widehat{\gamma}_5^{(\nu)}\hat{S}'_3)^2+\tilde{b}\Tr (\hat{S}'_3)^2]}
\end{split}
\end{equation}
with
\begin{equation}
\begin{split}
\tilde{a}=&\frac{\hat{N}^2((4\xi q)^2-z^2)^2}{16d_{\mathcal{G}}\kappa^2}\frac{(4\xi q)^2 \Tr(\hat{S}\widehat{\gamma}_5^{(N)})^2}{(4\xi q)^4\left(\Tr(\hat{S}\widehat{\gamma}_5^{(N)})^2\right)^2-z^4(\Tr\hat{S}^2)^2},\\
\tilde{b}=&\frac{\hat{N}^2((4\xi q)^2-z^2)^2}{16d_{\mathcal{G}}\kappa^2}\frac{z^2\Tr\hat{S}^2}{(4\xi q)^4\left(\Tr(\hat{S}\widehat{\gamma}_5^{(N)})^2\right)^2-z^4(\Tr\hat{S}^2)^2}
\end{split}
\end{equation}
To find the second line, we have shifted as follows
\begin{equation}
\hat{H}\to \hat{H}+2i \tilde{a}\widehat{\gamma}_5^{(\nu)}\hat{S}'_3\widehat{\gamma}_5^{(\nu)}-2i\tilde{b}\hat{S}'_3
\end{equation}
in the integrand in the denominator, which canceled the coupling term between $H$ and $\hat{S}'_3$.

To bring the result~\eqref{intermediate} into the well-known form of an elliptic Ginibre ensemble~\cite{akemann_baik_francesco}, we define the ellipticity
\begin{equation}
\begin{split}
\tau=&\frac{z^2\Tr\hat{S}^2}{(4\xi q)^2 \Tr(\hat{S}\widehat{\gamma}_5^{(N)})^2}=\frac{16N^2z^2}{(d_{\mathcal{G}}+\sqrt{d_{\mathcal{G}}^2+16N^2z^2})^2}\frac{{\rm Re}[\Tr S^2]}{ \Tr SS^\dagger}\ \in]-1,1[
\end{split}
\end{equation}
and fix the width of the ensemble
\begin{equation}
\kappa^2=\frac{\hat{N}^2((4\xi q)^2-z^2)^2}{4d_{\mathcal{G}}(4\xi q)^2 \Tr(\hat{S}\widehat{\gamma}_5^{(N)})^2}=\frac{d_{\mathcal{G}}}{\Tr SS^\dagger}.
\end{equation}
Then, the distribution simplifies to
\begin{equation}\label{Eq:EllipticGinibre}
p(S'_3)=\frac{\exp[-\frac{1}{1-\tau^2}\Tr S'_3{S'_3}^\dagger+\frac{\tau}{1-\tau^2}{\rm Re}[{\Tr}\,{ S'_3 }^2]]}{\int_{\mathcal{M}}dS'_3 \exp[-\frac{1}{1-\tau^2}\Tr S'_3{S'_3}^\dagger+\frac{\tau}{1-\tau^2}{\rm Re}[{\Tr}\,{S'_3}^2]]}.
\end{equation}
We would again like to emphasize that this formula holds for all non-Hermitian  symmetry classes. The subtle differences are encoded in the specific structure of $S'_3$. In all cases the distribution again has the form of a matrix-valued central limit theorem. 

In the perturbation $P_{\rm c}$ we can reintroduce the width $\kappa$ and the complex phase $e^{i\varphi}$ as well as the coupling constant $\alpha$ so that we finally arrive at the limiting perturbation $P_{\rm c}=\alpha e^{i\varphi}\kappa S'_3$ where $S'_3$ is distributed along~\eqref{Eq:EllipticGinibre}. The ellipticity vanishes when $z$ or ${\rm Re}[\Tr S^2]$ does,  while the spectrum becomes quasi 1-dimensional when $\tau\to\pm1$, this means we need both $|z|\to\infty$ and that ${\rm Re}[\Tr S^2]=\pm\Tr SS^\dagger$. The latter is only achieved when $S$ is almost Hermitian or anti-Hermitian in the large $N$ limit otherwise the ensemble only tends to an elliptic Ginibre ensemble with a fixed eccentricity.

The spectral density of \eqref{Eq:EllipticGinibre} for the case of $\mathcal{M}={\rm Gl}(N,\mathbb{C})$ can be found in Eq.~\eqref{one_point_corr}.


\section{Numerical Analysis}
\label{sec:numerical}

Below we {verify numerically that in the neighborhood of the real and imaginary axis of the eigenvalue density of the broadened zero modes follow the form given by the elliptic Ginibre ensemble. We begin} with the chiral ensemble,
\begin{align}
	K^{(N)}=
    	\begin{pmatrix}
      	0 & M \\
      	M^{\dagger} & 0
    	\end{pmatrix}
    	+\alpha_{R}U_{R}S_{R}U_{R}^{\dagger}+ i \alpha_{I}U_{I}S_{I}U_{I}^{\dagger}.
\end{align}
In this case, the distribution of the smallest eigenvalues is compared to the one-point correlation function (\ref{one_point_corr}) with the factor $\tau$ determined by the standard deviations found in Section \ref{sec:dist}. 

\begin{figure}[t]
\includegraphics[width=80mm]{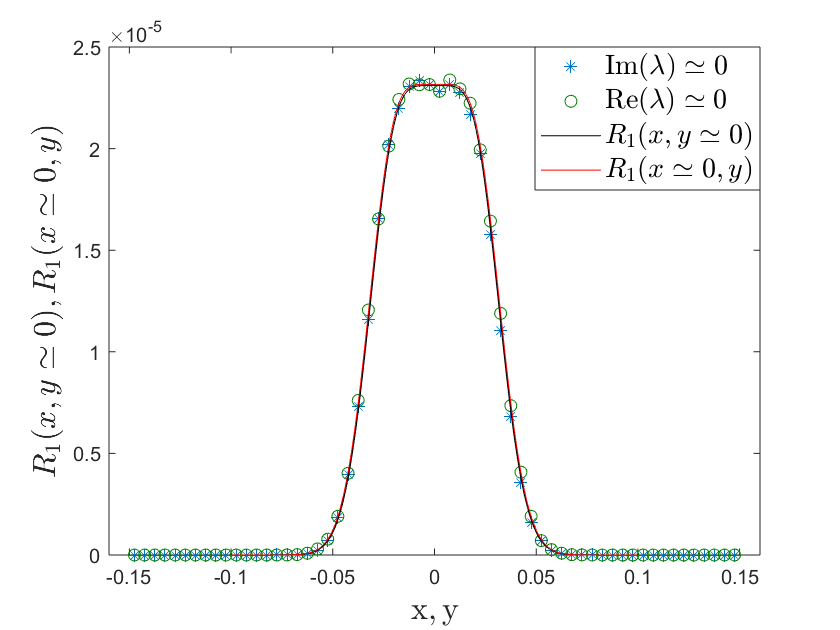}    	\includegraphics[width=80mm]{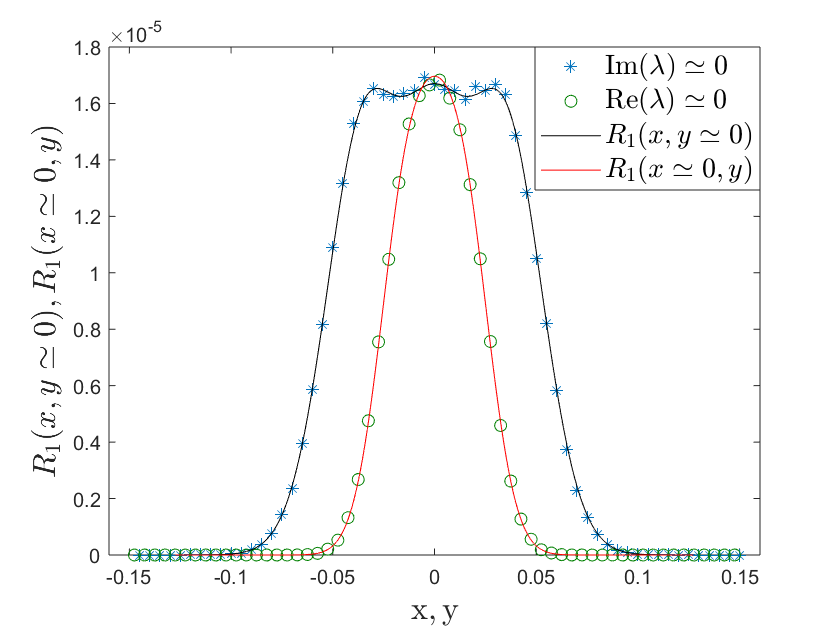}	
    \caption{The eigenvalue distribution of the chiral ensemble. Shown are the distributions of the smallest eigenvalues in a narrow bin either along the real- or the imaginary-axis. In the first case ({\bf left}) the perturbation is of equal magnitude in the real and imaginary direction, $\alpha_{R}=\alpha_{I}=0.01$ while in the second ({\bf right}) the perturbations are of different magnitude, $\alpha_{R}=0.015$ and $\alpha_{I}=0.01$. 
    In both figures the matrix dimension is $N=2n+\nu$ with $n=40$ and the number of zero modes is $\nu =3$. The numerical points are obtained from an ensemble of $10^6$ matrices. The eigenvalue density in the neighbourhood of the real and the imaginary axis follow the parameter free theoretical prediction given by respectively the circular and the elliptic Ginibre ensemble.}
    \label{fig:ensemble1_circular}
\end{figure}

The average is performed over the unitary matrices $U_{R}$ and $U_{I}$. The Hermitian matrices, $S_{R}$ and $S_{I}$, are kept fixed (the i.i.d. elements drawn from a normal distribution and the matrices are then diagonalised). The matrix $M$ is generically complex and of dimension $(n+\nu)\times n$, such that the full initial matrix is of dimension $N=2n+\nu$. The perturbation matrices are complex too, and have no further symmetries. Figure \ref{fig:ensemble1_circular} shows the results for $n=40$, $\nu=3$ and an ensemble of size $10^6$. Shown is the smallest eigenvalues in a narrow bin along the real and imaginary axis. Left panel shows the symmetric case $\alpha_{R}=\alpha_{I}=0.01$ while  the right panel shows $\alpha_{R}=0.015$ and $\alpha_{I}=0.01$. Also shown along with the numerical points are the parameter free analytic predictions and we observe complete agreement.

\begin{figure}[t]
    	\includegraphics[width=80mm]{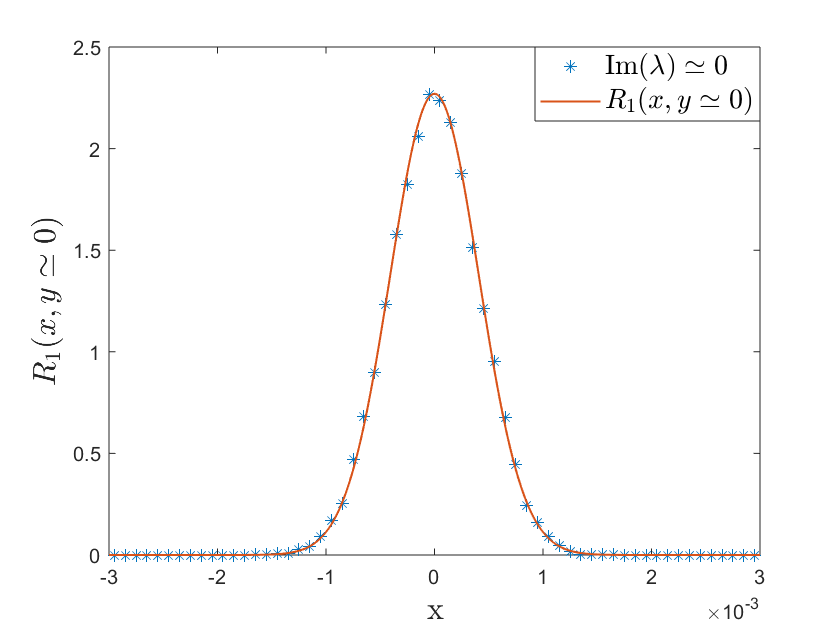}
    	\includegraphics[width=80mm]{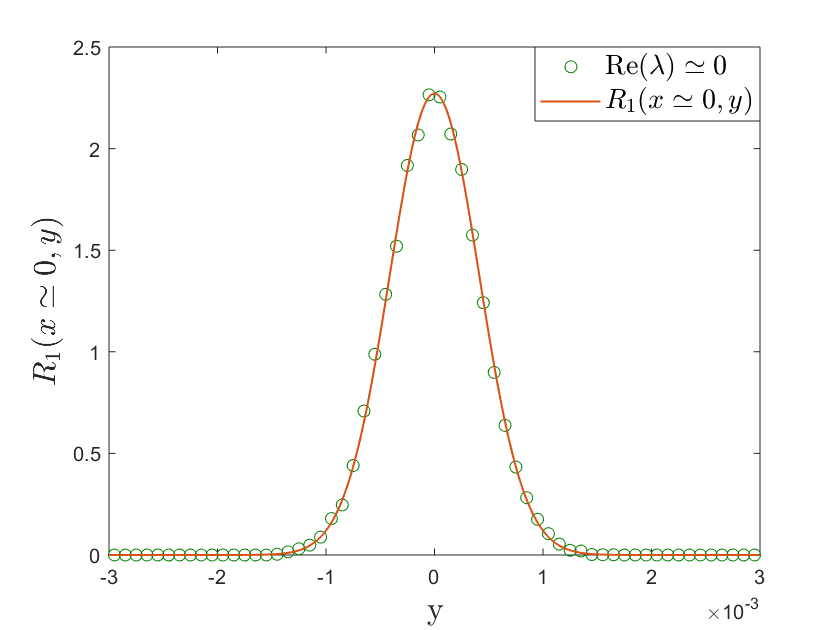}
    \caption{The eigenvalue distribution of the Majorana ensemble. As for the chiral ensemble we plot the smallest eigenvalues in a narrow bin close to respectively the real- ({\bf left}) or the imaginary-axis ({\bf right}). In both figures the numerical results are for matrix dimension $N=4n+2\nu$ with $n=20$ and $\nu =1$ and ensemble size $10^5$. The perturbation is of equal magnitude in the real and imaginary direction, $\alpha_{R}=\alpha_{I}=0.001$.}
    \label{fig:ensemble2}
\end{figure}

The second ensemble considered is the Majorana ensemble,
\begin{align}
  	K^{(N)}=
    	\begin{pmatrix}
      	 i M & 0\\
      	0 & - i M
    	\end{pmatrix}
    	+\alpha_{R}O_{R}
    	\begin{pmatrix}
      	0 &  i W_{R}\\
      	- i W_{R}^{T} & 0
    	\end{pmatrix}
    	O_{R}^{T}+ i \alpha_{I}O_{I}
    	\begin{pmatrix}
      	0 &  i W_{I}\\
      	- i W_{I}^{T} & 0
    	\end{pmatrix}
    	O_{I}^{T}, && M=-M^{T},
\end{align}
where, again, the only average performed is over the orthogonal transformations  $O_{R}$ and $O_{I}$. The elements of the matrices $M$ and $W$ are drawn only once as i.i.d. from a uniform distribution on the interval $[-1,1]$ and are then kept fixed. $M$ and $W$ are real and of dimension $2n+\nu$, and $M$ is antisymmetric. In figure \ref{fig:ensemble2} the broadening of the zero modes are compared to Gaussian distributions, see equation (\ref{Majorana_gauss}), in both the real and imaginary direction. Again complete agreement is found with the parameter free prediction.

\section{Bounds and Generalizations}
\label{sec:bounds}

It is natural to ask the question how universal the result \eqref{eq:GaussianS3} is. The main possible generalization is either a deformation of the Haar measure of $\hat{U}$ or {restricting to a subset of the original group}. Both kinds of deformations can be implemented by a distribution $g(\hat{U})$, in particular we consider the measure $g(\hat{U})d\tilde{\mu}(\hat{U})$ instead of the measure~\eqref{measure}. The impact of this deformation in the calculation appears when we want to integrate over the Gaussian, see~\eqref{Gauss-int}. This integral has now $g(\hat{U})$ as a prefactor,
\begin{equation}
\begin{split}
&\hat{g}(\eins_{\hat{N}}\otimes(\eins_{\hat{N}}-i\hat{Q})+i\frac{\kappa}{\xi}\hat{\Pi}_2^\dagger\widehat{\gamma}_5^{(\nu)}H\hat{\Pi}_2\otimes \hat{S}\widehat{\gamma}_5^{(N)}-\frac{z}{4\xi}\widehat{\gamma}_5^{(N)}\otimes\widehat{\gamma}_5^{(N)})\\
=&\frac{\int_{\mathcal{G}}d\hat{U} g(\hat{U})\exp[\frac{z}{4}\Tr[\hat{U}\widehat{\gamma}_5^{(N)}\hat{U}^\dagger\widehat{\gamma}_5^{(N)}]-i\kappa\Tr H\hat{\Pi}_2 \hat{U}\hat{S}\widehat{\gamma}_5^{(N)}\hat{U}^\dagger\hat{\Pi}_2^\dagger\widehat{\gamma}_5^{(\nu)}-\xi\Tr\hat{U}\hat{U}^\dagger(\eins_{\hat{N}}-i\hat{Q})]}{\int_{\mathcal{G}}d\hat{U}\exp[-\xi\Tr\hat{U}\hat{U}^\dagger]}.
\end{split}
\end{equation}
We can still exploit the factorization
\begin{equation}\label{trafo}
\begin{split}
&\hat{g}(\eins_{\hat{N}}\otimes(\eins_{\hat{N}}-i\hat{Q})+i\frac{\kappa}{\xi}\hat{\Pi}_2^\dagger\widehat{\gamma}_5^{(\nu)}H\hat{\Pi}_2\otimes \hat{S}\widehat{\gamma}_5^{(N)}-\frac{z}{4\xi}\widehat{\gamma}_5^{(N)}\otimes\widehat{\gamma}_5^{(N)})\\
=&{\det}^{-d_{\mathcal{G}}/(4\hat{N})}\left[(\eins_{\hat{N}}-i\hat{Q})^2-\frac{z^2}{16\xi^2}\eins_{\hat{N}}\right]\\
&\times \hat{g}\left(\eins_{\hat{N}^2}+i\frac{\kappa}{\xi}\hat{\Pi}_2^\dagger\widehat{\gamma}_5^{(\nu)}H\hat{\Pi}_2\otimes \hat{S}\widehat{\gamma}_5^{(N)}\left[\eins_{\hat{N}}\otimes(\eins_{\hat{N}}-i\hat{Q})-\frac{z}{4\xi}\widehat{\gamma}_5^{(N)}\otimes\widehat{\gamma}_5^{(N)}\right]^{-1}\right).
\end{split}
\end{equation}
This result can be achieved by rescaling $\hat{U}$ as follows
\begin{equation}
\begin{split}
\hat{U}_{kl} \rightarrow \sum_{k',l'=1}^{\hat{N}}\hat{U}_{k'l'}\left\{\left(\eins_{\hat{N}}\otimes(\eins_{\hat{N}}-i\hat{Q})-\frac{z}{4\xi}\widehat{\gamma}_5^{(N)}\otimes\widehat{\gamma}_5^{(N)}\right)^{-1/2}\right\}_{k'k,l'l}.
\end{split}
\end{equation}

In~\eqref{trafo}, we see a sufficient condition for not influencing the result apart from a rescaling of the width $\kappa$. We need to assume that the transform $\hat{g}$ of the function $g(\hat{U})$ remains finite and differentiable in the vicinity of the identity matrix $\eins_{\hat{N}^2}$. The dependence on $N$ can be weak as it is in the case for the determinant in our computation above. Then, the Taylor expansion of ${\rm ln}(\hat{g})$ about  $\eins_{\hat{N}^2}$ can be carried out up to the second order. The $N$-dependence of higher order coefficients should not interfere with the behavior of the ratio $1/q^{(N)}\to0$, see~\eqref{ratio-limit2}, so that those terms still vanish in this limit. As a by-product of this calculation, we can read off the rescaling of the width $\kappa$ which is the factor $1/\sqrt{\hat{g}(\eins_{\hat{N}^2})}$.

Though the discussion above has been only for the second model~\eqref{eq:Model_c}, a similar analysis can be expected for the first model~\eqref{eq:Model_unc}, as well. As already mentioned, both computations follow along the same steps.


\section{Conclusion}
\label{sec:conclusion}

The remarkable fact that perturbed zero modes distribute themselves according to finite size Gaussian random matrix theory has been shown to apply also for non-Hermitian perturbations. In the non-Hermitian case, the distribution of the perturbed zero modes follows the one point function of the elliptic Ginibre ensemble.  
At first, it may appear highly surprising that finite size Gaussian random matrix theory can provide universal distributions for would-be zero modes. Usually random matrix theory only yields universal results for infinite size matrices and the Gaussian in the weight can be replaced by some other weight without altering the universal distribution as long as the support of the eigenvalue density is still a single cut. For the perturbed zero modes, however, the Gaussian weight in the universal distribution can not be changed. As shown, the Gaussian form results from a matrix version of the central limit theorem which applies to the finite size perturbation matrix as the size of the full Hamilton goes to infinity.  
We have derived explicit bounds where the results are applicable. In physical terms, the bounds correspond to the regime where first order degenerate perturbation theory dominates. 

As physical examples, we have shown that the broadened zero modes of both a chiral matrix, with the same symmetry properties as the Dirac operator of QCD, and of a matrix modeling topological superconductors with Majorana particles, follow an elliptic Ginibre ensemble, when perturbed by a non-Hermitian perturbation. 

We have demonstrated the universality of the results by considering two different forms of the non-Hermitian perturbation, both leading to the same universal form for the perturbed zero modes. Furthermore we have discussed the stability of the results when restricting the manifold over which the average is performed. It would be most interesting to generalize these universality considerations.

\begin{acknowledgments}

We would like to thank Karsten Flensberg, Esben Hansen, and the participants the workshop 'RMT in Sub-Atomic Physics and Beyond' held at ECT* during August 2019 for discussions. Supported by the German research council DFG through International Research Training Group 2235 Bielefeld-Seoul ``Searching for the regular in the irregular: Analysis of singular and random systems.'' (A.M.)

\end{acknowledgments}


\end{document}